\documentclass{pasj00} 
\draft
\begin{document}
\SetRunningHead{K.~Koyama \etal}{Discoveries of Diffuse Iron Line Sources from the Sgr B Region}
\Received{2006/07/05}  
\Accepted{2006/08/**}  

\title{Discoveries of Diffuse Iron Line Sources from the Sgr B Region}

\author{Katsuji \textsc{Koyama},  Tatsuya \textsc{Inui}, Yoshiaki {\sc Hyodo}, Hironori {\sc Matsumoto} and  Takeshi Go {\sc Tsuru}}

\affil{Department of Physics, Graduate School of Science, Kyoto University, Sakyo-ku, Kyoto 606-8502}
\email{koyama@cr.scphys.kyoto-u.ac.jp, inuit@cr.scphys.kyoto-u.ac.jp}

\author{Yoshitomo   \textsc{Maeda}} 

\affil{Institute of Space and Astronautical Science, JAXA, Sagamihara, Kanagawa, 229-8510}

\author{Hiroshi \textsc{Murakami}}
\affil{PLAIN center, ISAS/JAXA, 3-1-1 Yoshinodai, Sagamihara, 
Kanagawa 229-8510}

\author{Shigeo \textsc{Yamauchi}}

\affil{Faculty of Humanities and Social Sciences, Iwate University, 3-18-34 Ueda, Morioka, Iwate 020-8550}

\author{Steven E. \textsc{Kissel}}
\affil{Kavli Institute for Astrophysics and Space Research, Massachusetts Institute of Technology, 
Cambridge, \\
MA 02139, USA}

\author{Kai-Wing \textsc{Chan}  and Yang \textsc{Soong}}
\affil{Code 662. NASA/GSFC, Greenbelt, MD 20771, USA}


\maketitle

\begin{abstract}
The radio complex Sgr B region is observed with the X-Ray Imaging Spectrometers (XIS) on board Suzaku. This region exhibits diffuse iron lines at 6.4, 6.7 and 6.9~keV, which are K$\alpha$ lines of Fe \emissiontype{I} (neutral iron), Fe\emissiontype{XXV} (He-like iron) and Fe\emissiontype{XXVI} (H-like iron), respectively. The high energy resolving power of the XIS provides the separate maps of the K-shell transition lines from Fe\emissiontype{I} (6.4~keV) and Fe\emissiontype{XXV} (6.7~keV). Although the 6.7~keV line is smoothly distributed over the Sgr B region, a local excess is found near at $(l, b) = (\timeform {0D.61}, \timeform{0D.01})$, possibly a new SNR. The plasma temperature is \textit{kT} $\sim$3~keV and the age is estimated to be around several$\times10^{3}$ years. The 6.4~keV image is clumpy with local excesses nearby Sgr B2 and at $(l, b) = (\timeform{0D.74}, -\timeform{0D.09})$. Like Sgr B2, this excess may be another candidate of an X-ray reflection nebula (XRN).
\end{abstract}

\section{Introduction} 
Sgr B is a molecular complex consisting of Sgr B1 and B2. In particular, Sgr B2 is the most massive giant molecular cloud with ultra compact H\emissiontype{II} regions (UCH\emissiontype{II}) \citep{Gaume1995} and many maser sources near the cloud center \citep{Mehringer1997}. These are the hints of high mass (HM) zero age main sequence (ZAMS) stars or young stellar objects (YSOs). However, extremely high absorption toward the cloud center ($N_{\rm H} \ge 10^{24}~{\rm cm}^{-2}$, $A_{\rm V}$= a few 100) prevents to detect any stars in optical or even in the infrared bands. \citet{Takagi2002} have discovered many compact X-ray sources in the cloud center with Chandra. The X-ray fluxes and spectra indicate that these are likely HM YSOs. Since HM stars evolve very rapidly and finally undergo  supernova explosions, it will be reasonable to expect young SNRs near the Sgr B2 cloud. In facts, \citet{Senda2002} have discovered a peculiar SNR candidate with Chandra. 

Sgr B2 is also a strong 6.4~keV line source. \citet{Koyama1996} and \citet{Murakami2001} proposed that Sgr B2 is an X-ray reflection nebula (XRN), irradiated by the Galactic center (GC) source Sgr A$^{*}$. Sgr A$^*$ was  then  thought to be X-ray bright about 300 years ago, the light traveling time between Sgr B2 and Sgr A$^{*}$. In this XRN scenario, it is likely that other XRNe will be found in the molecular complex Sgr B. The Suzaku observation on the Sgr B region is intended to discover new SNRs and XRNe. This paper reports the first results of the Suzaku observation.

\section{Observation and Data Processing} 

\subsection{Data Collection} 

The Sgr B region was observed with the XIS on 10-12 October 2005. The XIS consists of four sets of X-ray CCD camera systems (XIS0, 1, 2, and 3) placed on the focal planes of four X-Ray  Telescopes (XRT) on board the Suzaku satellite. XIS0, 2 and 3 have front-illuminated (FI) CCDs, while XIS1 has a back-illuminated (BI) CCD. The detail descriptions of the Suzaku satellite, the XRT and  the XIS are found in \citet{Mitsuda2006}, \citet{Serlemitosos2006} and Koyama et al. (2006a), respectively. 
The XIS observation was made with the normal mode. The effective exposure time after removing the epoch of low earth elevation angle ( ELV $\le$5$^\circ$ ) and the South Atlantic Anomaly was about 89~ksec. 

\subsection{The Gain Tuning} 
In a quick look of the spectrum, we found strong  lines at $\sim$6.7~keV and $\sim$6.4~keV, in everywhere in the CCD imaging area (IA), which may be due to the largely extended Galactic center diffuse X-ray emission (GCDX). These lines are most likely K{$\alpha$} lines of Fe\emissiontype{XXV} (6.7~keV) and Fe\emissiontype{I} (6.4~keV). Using the center energies of the two strong lines, we made fine correction  of the CTI (Charge Transfer Inefficiency), and fine gain tuning in XIS-to-XIS and segment-to-segment levels (relative gain tuning). Then the absolute gain tuning is made using the $^{55}$Fe calibration sources irradiating the CCD corners. Details of this procedure and high capability are demonstrated  in Koyama et al. (2006b).

\subsection{The Position Tuning} 

\begin{figure}
  \begin{center}
    \FigureFile(80mm,80mm){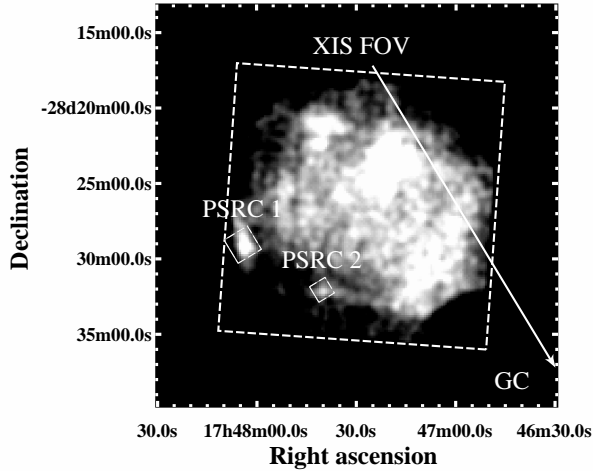}
  \end{center}
  \caption{The X-ray image of the Sgr B region in the 2--10 keV band. All the four XIS data are co-added. The dotted square is the XIS field of view (FOV).}
  \label{fig:suzaku2-10keVimg}
\end{figure}

After the CTI correction and fine gain tuning, we add all the XIS data and made a composite image
of the Sgr B region in the 2--10 keV band (figure \ref{fig:suzaku2-10keVimg}). The diffuse enhancement in the northwest corresponds to 
the Sgr B2 complex. Other than this, we found  two point sources at the southeast edge of the XIS field of view (FOV).  We made the radial profiles of these 
two sources and determined the peak positions to be $(l,~~b) = (\timeform{0D.5762},~ -\timeform{0D.1736})$ and 
$(\timeform{0D.6626},~ -\timeform{0D.2225})$ in the nominal Suzaku coordinate. We search for the Chandra Galactic center 
survey map, and found possible counterparts, CXO~J174741$-$283213 and  CXO~J174804.8$-$282919 \citep{Muno2003}. The Galactic coordinates of these sources
are $(l,  b) = (\timeform{0D.5762}, -\timeform{0D.1796})$ and $(\timeform{0D.6625}, -\timeform{0D.2289}$).  Therefore the Suzaku nominal 
coordinate is systematically shifted by $(\Delta l, ~\Delta b) = 
(-\timeform{0D.0001}, ~-\timeform{0D.0062})$ from the Chandra coordinate. Since the aspect solution of Chandra is very accurate within 
sub-arcsec, we made fine tuning of the Suzaku position by shifting $(-\timeform{0D.0001},~ -\timeform{0D.0062})$ in 
the $(l,~ b)$ coordinate. Hereafter we use this re-registered coordinate.

\vspace*{1 cm}

\section{Results and Discussions} 
\subsection{The Overall Features} 

\begin{figure}
  \begin{center}
    \FigureFile(80mm,80mm){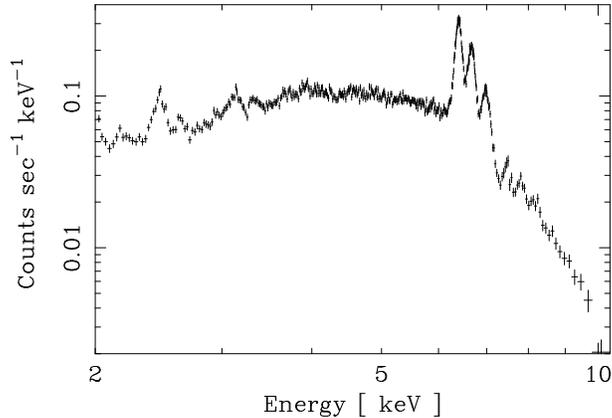}
  \end{center}
  \caption{The X-ray spectra from the full FOV of the XIS, but the CCD corners irradiated by the build-in calibration sources are excluded. All the four XIS data are co-added}
  \label{fig:sgrb-overall}
\end{figure}

The X-ray spectrum of all the Sgr B region is given in figure \ref{fig:sgrb-overall}. The spectra of the four XISs (XIS0-XIS3) are co-added and the night earth spectrum (non X-ray background; here, NXBG) is subtracted. With the superior energy resolution of the XIS for diffuse sources, we can clearly resolve the 6.4~keV, 6.7~keV and 6.9~keV lines. These are K$\alpha$ lines from  neutral Fe\emissiontype{I}, He-like Fe\emissiontype{XXV} and hydrogenic Fe\emissiontype{XXVI}. 
The 6.9~keV line may contain a small fraction of K$\beta$ line of Fe\emissiontype{I} (7.07 keV).
Weak lines seen above ~7 keV are 
K$\alpha$ of Ni\emissiontype{I}(at $\sim$ 7.5 keV), 
K$\alpha$ of Ni\emissiontype{XXVII} + K$\beta$ of Fe\emissiontype{XXV} (at $\sim$7.8--7.9 keV) and 
K$\beta$ of Fe\emissiontype{XXVI} + K$\gamma$ of Fe\emissiontype{XXV} (at $\sim$8.2--8.3 keV).

\subsection{Discovery of a new SNR} 
\vspace*{1 cm}
\begin{figure}
  \begin{center}
    \FigureFile(80mm, 10mm){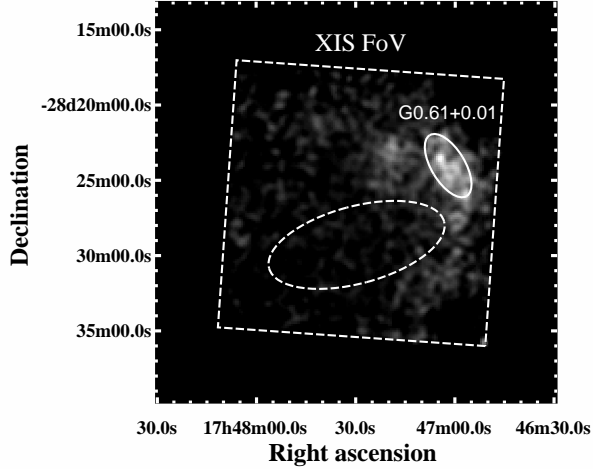}
  \end{center}
  \caption{The 6.7~keV line map (the 6.58--6.74 keV band map) showing bright spot at the northwest corner. The fluxes are normalized by the 6.7~keV flat-field image. The source and background regions are shown by the solid and dotted ellipses, respectively.}
  \label{fig:sgrb-6700img}
\end{figure}
We made a narrow band image of 6.7~keV (the 6.58--6.74 keV band) in figure \ref{fig:sgrb-6700img}. We see a clear 6.7~keV flux excess at the northwest corner. 
To confirm the 6.7 keV excess, we referred the archive data of Chandra (OBSID:944, effective exposure time was 99 ksec)
and XMM (OBSID: 0203930101, effective exposure time was 42~ksec). Since the energy resolution of the Chandra ACIS is limited 
to separate the 6.4 and 6.7 keV lines, it is unclear whether the 6.7 keV source is present or not. The XMM image in the 6.7 keV band shows a clear elongated structure near the same position. Accordingly the presence of the 6.7 keV line source is no doubt.
On the other hand, the continuum band image (e.g. 2-5 keV, or 2-8 keV bands) shows only a hint of enhancement, and show no clear structure.  This source is therefore very peculiar, which is dominant only in the 6.7 keV line.
Since the dominance of the 6.7 keV line suggests that the excess is a new SNR (see also below for the discussion), we designate this source as Suzaku~J1747.0$-$2824.5 (G0.61+0.01) from its center position.  

\begin{figure}
  \begin{center}
    \FigureFile(80mm,80mm){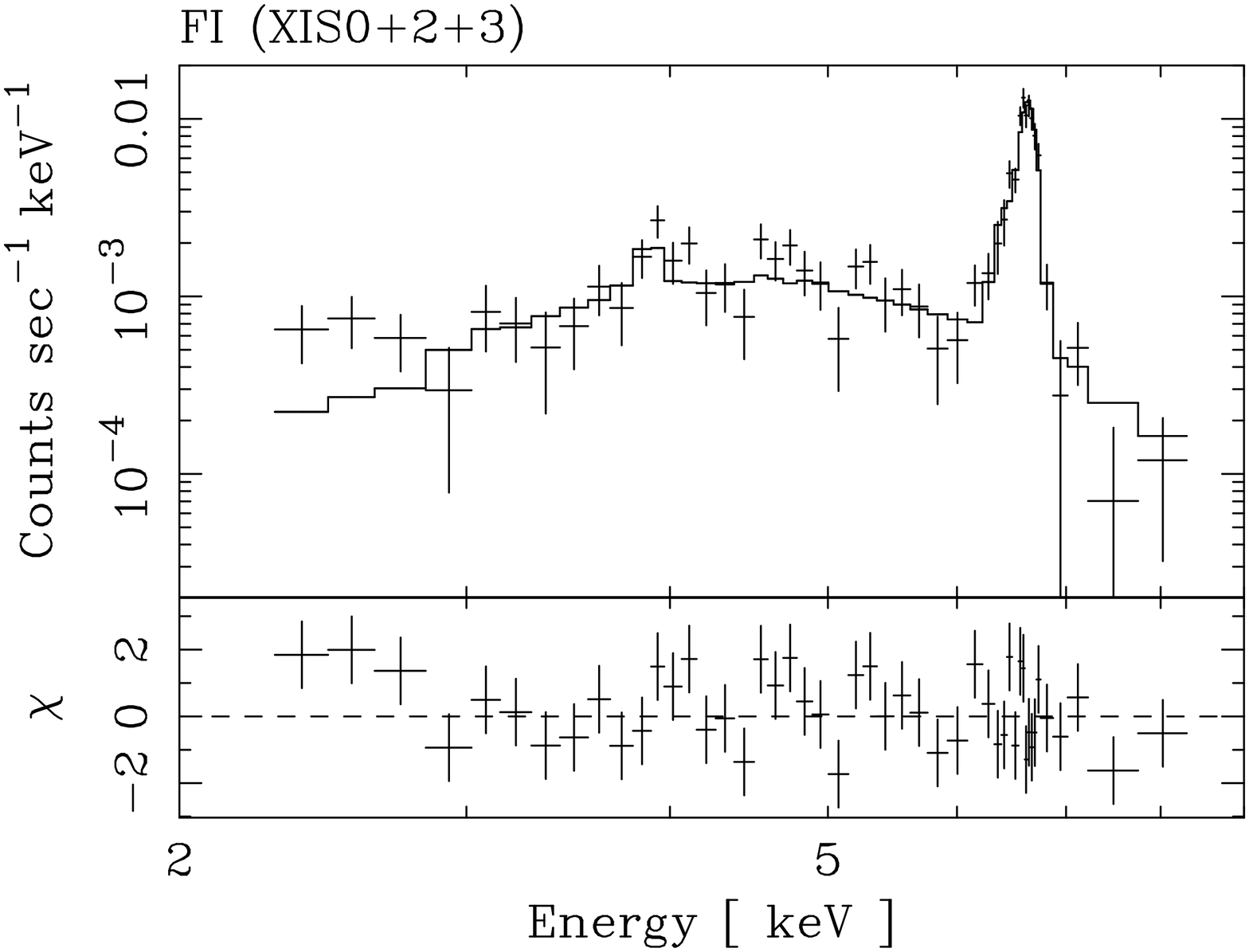}
    \FigureFile(80mm,80mm){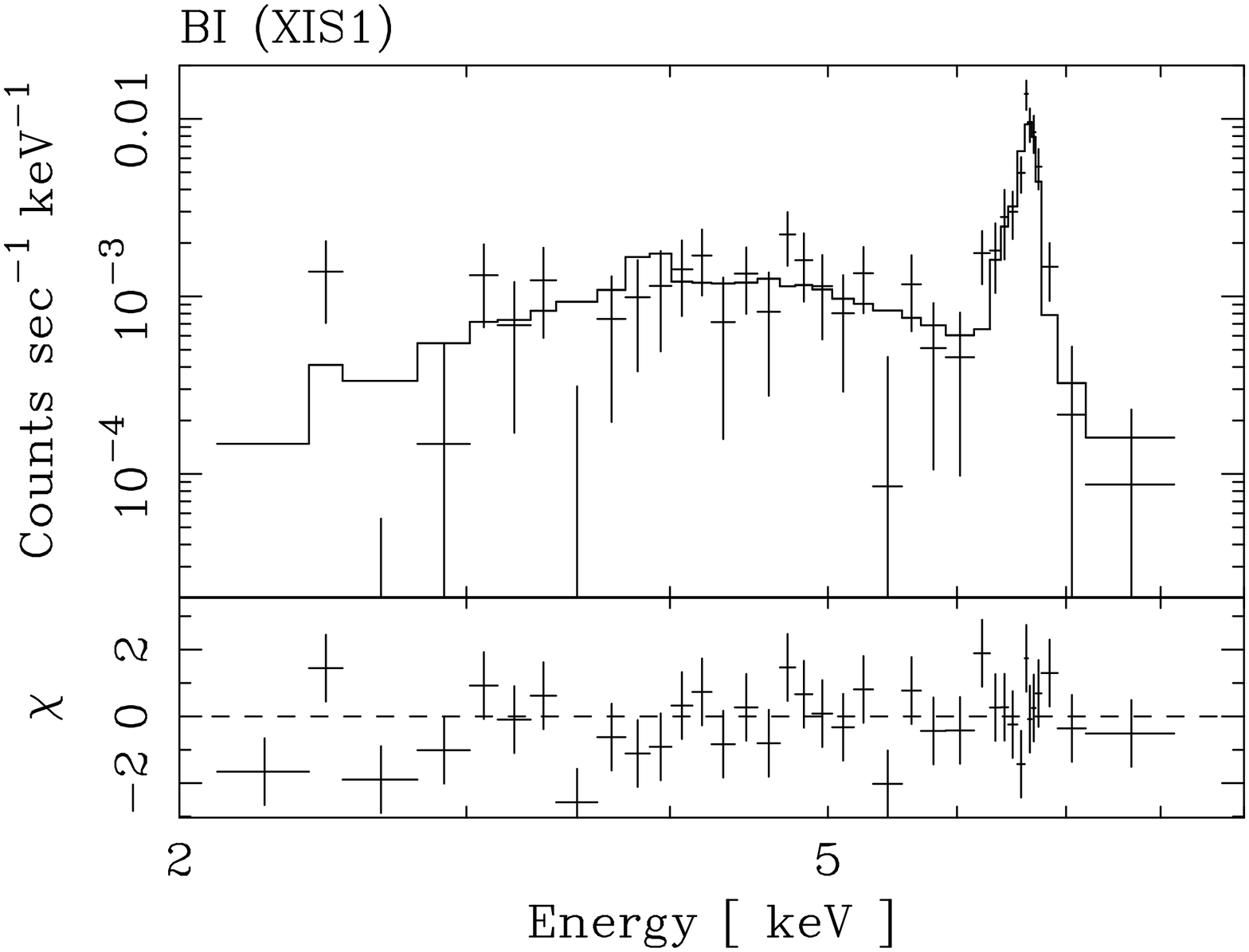}
  \end{center}
  \caption{Left: the X-ray spectrum of the sum of 3 FI CCDs (XIS0, 2 and 3) for a new SNR (G0.61+0.01) with the best-fit VPSHOCK model. Right: same as the left, but of the BI CCD (XIS1)}  \label{fig:sgrb-g061001-spec}
\end{figure}

We made the NXBG-subtracted spectra from the solid ellipse for both the three FI CCDs (XIS0, XIS2 and XIS3 are co-added) and BI CCD (XIS1). In these spectra, the cosmic X-ray  background (CXB) and GCDX  are still included.  We therefore made the NXBG-subtracted spectra from the dotted ellipse in figure \ref{fig:sgrb-6700img},
and subtract this local background (CXB + GCDX) from the source spectrum.
All the spectra  have  been corrected for the vignetting at 6.7~keV. 
The results are given in figure \ref{fig:sgrb-g061001-spec}, for the FIs and  BI, separately.  We see a pronounced peak at 6.7~keV, but no 6.9 keV line.
The 6.7 line shape is asymmetric with a tail at lower energy. 
In order to verify the line structure, we drive fluxes 
of the 6.4, 6.7 and 6.9 keV lines ( K$\alpha$ line of Fe \emissiontype{I}, \emissiontype{XXV} and  \emissiontype{XXVI}) in the source
and the background regions,
applying a phenomenological model (a bremsstrahlung continuum and many Gaussian lines) in the raw data (no background subtraction). 
The resulting  6.4,  6.7 and 6.9 keV line fluxes are  2.22, 5.17 and 0.48 for the G0.61+0.01 (source) region, and  0.61, 0.68  and 0.30 for the background region, 
where the flux unit is  10$^{-6}$ photons~cm$^{-2}$~s$^{-1}$ ~arcmin$^{-2}$. In contrast to the 6.7 keV line, we see no large excess in the 6.9 keV line from the source region compared to the background region. Thus we confirm that G0.61+0.01 emits
strong  6.7 keV line, but very weak 6.9 keV line.
The  small excess of the  6.4 keV line makes the low energy  tail in the 6.7 keV line.

\begin{table}
  \caption{The best-fit parameters for G0.61+0.01 with the VPSHOCK model
plus
  two emission lines}
  \label{tab:g_fit}
  \begin{center}
    \begin{tabular}{lc}
      \hline\hline
      Parameter & Value \\
      \hline
      $N_{\rm H}$ ($10^{23}$~H~cm$^{-2}$) & $1.6_{-0.4}^{+0.7}$ \\
      $kT$ (keV) & $3.2_{-0.9}^{+2.3}$ \\
      $n_{\rm e}t$ ($10^{11}$ cm$^{-3}$ s) & $1.9_{-0.8}^{+4.7}$ \\
      \multicolumn{2}{l}{Abundances\footnotemark[$*$]}\\
      Ca & $3.5_{-2.4}^{+3.1}$ \\
      Fe & $5.1_{-1.1}^{+1.2}$ \\
      \multicolumn{2}{l}{Neutral iron lines\footnotemark[$a$]}\\
      $I_{6.40}$ ($10^{-6}$ photons cm$^{-2}$ s$^{-1}$) & $5.1_{-2.5}^{+
2.4}$ \\
      $I_{7.06}$ ($10^{-6}$ photons cm$^{-2}$ s$^{-1}$) & 0.6 \\
      \multicolumn{2}{l}{Flux and Luminosity}\\
      $F_{\rm 2-10}$\footnotemark[$\dagger$] ($10^{-13}$ ergs cm$^{-2}$
s$^{-1}$) & $7.5_{-2.2}^{+1.1}$ \\
      $L_{\rm 2-10}$\footnotemark[$\ddagger$] ($10^{34}$ ergs s$^{-1}$)
& $1.5_{-0.2}^{+0.1}$ \\
      \hline
      $\chi^2$/dof & 99.0/78\\
      \hline
      \multicolumn{2}{@{}l@{}}{\hbox to 0pt{\parbox{80mm}{\footnotesize
        Note---The uncertainties indicate the 90\% confidence limit.
        \par\noindent
        \footnotemark[$*$] The elements which are not listed below are
fixed at
    1.0 (solar ratio).
        \par\noindent
        \footnotemark[$a$] The line energy of K$\alpha$ and K$\beta$ is
fixed at the theoretical value (6.40 and 7.06~keV, respectively; Kaastra and Mewe 1993) and the
intensity of K$\beta$ is fixed at 12.5\% (Kaastra and Mewe 1993) of that of K$\alpha$.
        \par\noindent
        \footnotemark[$\dagger$] Observed flux in the 2.0--10.0~keV band.
        \par\noindent
        \footnotemark[$\ddagger$] Absorption corrected luminosity in the
2.0--10.0~keV band.
      }\hss}}
    \end{tabular}
  \end{center}
\end{table}

The FI and BI spectra are simultaneously fitted with a plane parallel shock model (VPSHOCK in the XSPEC package) adding two Gaussian lines at 6.4~keV and 7.06~keV. These two lines represent the K$\alpha$ and K$\beta$ lines of Fe \emissiontype{I}, where the flux of latter line is fixed at 12.5\% of the former (Kaastra and Mewe 1993). The best-fit results and parameters are shown in figure \ref{fig:sgrb-g061001-spec} and table \ref{tab:g_fit}.
Although we detected the 6.4 keV line from G0.61+0.01,  it is very difficult to judge whether this line is really attributable to  G0.61+0.01, due to spilled-over flux from the adjacent source Sgr B2,  or due to a fluctuation of a larger scale structure in the 6.4~keV line.  As for the last possibility,  we see a large scale 6.4~keV enhancement in the northwest compared to 
the background region in the southeast (see figure \ref{fig:sgrb-6400img}).  In any case,  we ignore this line in the discussion of G0.61+0.01 because the 6.4~keV line flux is only 3\% of that in Sgr B2 (see tables 1 and  3). 
Since the Suzaku spatial resolution is not good enough, there may be possible contamination of unresolved point sources. To check this problem,
we searched for point sources using the Chandra archive data (OBSID: 944, ~99 ksec exposure time) and found no point source in the source region. On the other
hand,  in the background region, there are 48 point sources. The total flux in the 2-10 keV band is  5$\times10^{-13}$ ergs~cm$^{-2}$~s$^{-1}$, which is only $\sim$2\% of the CXB + GCDX flux of 2.6$\times10^{-11}$ ergs~cm$^{-2}$~s$^{-1}$, hence can be ignored in the present data analysis and discussion.

The best-fit temperature of $\sim$3~keV and overabundance of Fe are consistent with an ejecta of  an SNR and   
are similar to those found in the central region of Sgr A East, a young  SNR near at the GC.
The high temperature component of Sgr A  East is $kT\sim$4--6 keV (Sakano et al. 2004, Park et al. 2005, Koyama et al. 2006c),  and iron is overabundant by factor of 4--5 (Maeda et al. 2002, Sakano et al. 2004, Park et al. 2005, Koyama et al. 2006c).  Thus G0.61+0.01 is likely an ejecta dominant central region of an SNR.
We  note that Sgr A East has low temperature component of about 1 keV, while not in  G0.61+0.01. The absence of softer plasma may be due to the large absorption.
The $N_{\rm H}$ value of $1.6\times 10^{23}$~H~cm$^{-2}$ is larger than that of typical value to the GC ($6\times 10^{22}$~H~cm$^{-2}$ ) 
\citep{Sakano2002}. Therefore, G0.61+0.01 would be located behind or in the rim of the Sgr B2 cloud. Since G0.61+0.01 is located in the south of an expanding radio shell \citep{Oka1998}, which is probably interacting with the Sgr B2 cloud rim, we assume that the distance of G0.61+0.01 is the same as Sgr B2 and to be 8.5~kpc \citep{Reid1988}. Then the 2--10 keV band luminosity is estimated to be $1.5\times10^{34}$ ergs~s$^{-1}$, which is typical for an ejecta plasma of an  SNR.
The size of G0.61+0.01 (the solid ellipse in figure \ref{fig:sgrb-6700img}) is
~2.2$^\prime\times$4.8$^\prime$, which corresponds to ~5.5 pc~$\times$~12 pc  at a distance of 8.5 kpc.  
Assuming the plasma emission is due to a uniform density ellipsoid with the 3-axis radii of 2.7 pc, 2.7 pc 
and 6 pc, we estimate physical parameters of G0.61$+$0.01 ( table \ref{tab:g_physpar}). 
Although  the iron abundance is 3-4 times of the solar, total number and mass of 
protons are fur larger  than those  of irons.  We therefore assume that electron density ($n_{\rm e}$) is equal to the proton density ($ n_{\rm p}$), and that protons 
carry most of the plasma mass ($m_{\rm p}$).
Dividing the radius of the major axis (6 pc)  by the sound velocity of the 3.2~keV plasma ($v=1.4\times$10$^8$ cm~s$^{-1}$), we obtain the dynamical time scale ($t_{dyn}$) of $\sim$4$\times10^{3}$ years. 
If, instead, we  use the ionization parameter ($n_{\rm e}t$) and electron density ($n_{\rm e}$), then the  
ionization time scale ($t_{ioni}$) is estimated to be $\sim7\times10^{3}$ years. 
Since the source size of ~2.2$^\prime\times$4.8$^\prime$ is comparable to that of the half power diameter ($\sim2^\prime$), the real size of G0.61+0.01 must be smaller.
Therefore the quoted value of $t_{dyn}\sim4\times$10$^{3}$ years should be an upper limit, while that of  $t_{ioni}\sim7\times$10$^{3}$ years is 
a lower limit, because $n_{\rm e}$ is inversely proportional to the root square of the plasma volume.
Thus the age of G0.61+0.01 is probably around several$\times$10$^{3}$ years.

Another possibility is that  G0.61+0.01 comprises  a part of a larger SNR. Since G0.61+0.01 is found at the edge of the XIS field, other parts of a candidate  SNR may be out of the XIS field. In this scenario, G0.61+0.01 may be  a part of the expanding radio shell discovered by \citet{Oka1998}. The kinetic energy of the radio shell is a few of $10^{52}$ erg s$^{-1}$, within the range of single or multiple supernova explosions. Thus follow-up X-ray observations including this expanding radio shell is highly required.
\begin{table}
  \caption{The physical parameters of G0.61+0.01}
  \label{tab:g_physpar}
  \begin{center}
    \begin{tabular}{lc}
      \hline\hline
      Parameter & Value \\
      \hline
      EM\footnotemark[$a$] (cm$^{-3}$) & $1.4\times 10^{57}$ \\
      $n_{\rm e}$\footnotemark[$b$] (cm$^{-3}$) & $0.9$ \\
      $M$\footnotemark[$c$] (\MO) & 1.3 \\
      $E$\footnotemark[$d$] (ergs) & $2.4\times 10^{49}$ \\
      $t_{\rm dyn}$\footnotemark[$e$] (s) & $1.3\times10^{11}$ \\
      $t_{\rm ioni}$\footnotemark[$f$] (s) & $2.1\times10^{11}$ \\
      \hline
      \multicolumn{2}{@{}l@{}}{\hbox to 0pt{\parbox{60mm}{\footnotesize
        Note---The plasma is assumed to be a uniform density ellipsoid with the 3-axis radii of 2.7 pc, 2.7 pc and 6 pc (see text). 
        \par\noindent
        \footnotemark[$a$] Emission measure (EM) = $n_{\rm e} n_{\rm H}V$ = $n_{\rm e}^{2}V$,
    where $n_{\rm e}$ and $n_{\rm H}$ are  the electron and hydrogen  density and are assumed to be equal.
        \par\noindent
        \footnotemark[$b$] The electron density.
        \par\noindent
        \footnotemark[$c$] Total mass $(M) = n_{\rm e}m_{\rm p}V$, where $m_{\rm p}$ is the proton mass and $V$ is the plasma volume. 
    \par\noindent
        \footnotemark[$d$] Thermal energy $(E) = 3n_{\rm e}kTV$.
    \par\noindent
        \footnotemark[$e$] The dynamical time scale:  the radius of the major axis of the plasma ellipsoid divided by the sound velocity
of the $\sim$3 keV plasma. 
    \par\noindent
        \footnotemark[$f$] The ionization time scale: the ionization parameter (see table 1) divided by the electron density
}
\hss}}
    \end{tabular}
  \end{center}
\end{table}

\subsection{Discovery of a New XRN}

\begin{figure}
  \begin{center}
    \FigureFile(80mm,80mm){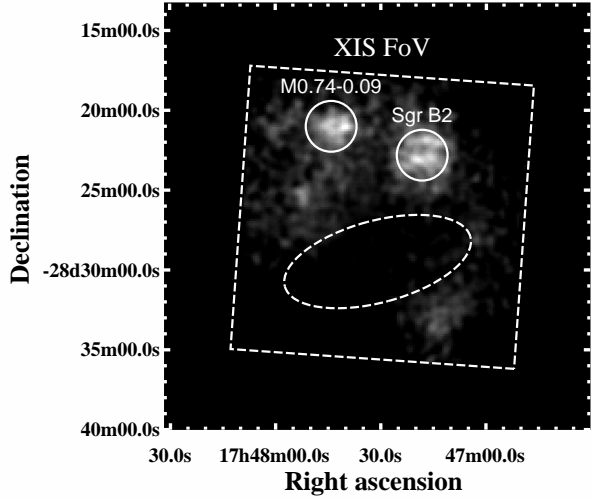}
  \end{center}
  \caption{The 6.4~keV line map (the 6.33--6.46 keV band map) showing bright spots at Sgr B2 and M0.74-0.09. 
  The fluxes are normalized by the 6.4~keV flat-field image.
  The sources and background regions are shown by the solid circles and dotted ellipse, respectively.}\label{fig:sgrb-6400img}
\end{figure}

We have made a narrow band image at 6.4~keV (the 6.33--6.46 keV band) in figure \ref{fig:sgrb-6400img}. We see two bright spots in the north. One is Sgr B2 which has been already found as a strong 6.4 keV source \citep{Koyama1996}, and the other is a newly discovered source. 
We again referred the same archive data of Chandra and XMM as the case of G0.61+0.01.  In the Chandra data, this excess is found near 
the edge of the ACIS  FOV. The XMM data show a clear excess near this source.  
The presence of the 6.4~keV line supports the presence of cool and dense gas clouds. We therefore designate this new source as Suzaku~J1747.7$-$2821.2 
(M0.74$-$0.09) from its peak position. 
We made X-ray spectra of Sgr B2 and M0.74$-$0.09 from the circles given in figure \ref{fig:sgrb-6400img}. 
The former is for comparison to the latter new source. The background spectrum is made from the dotted ellipse and subtracted in the same procedure as the case of G0.61+0.01.

\begin{figure}
  \begin{center}
    \FigureFile(80mm,80mm){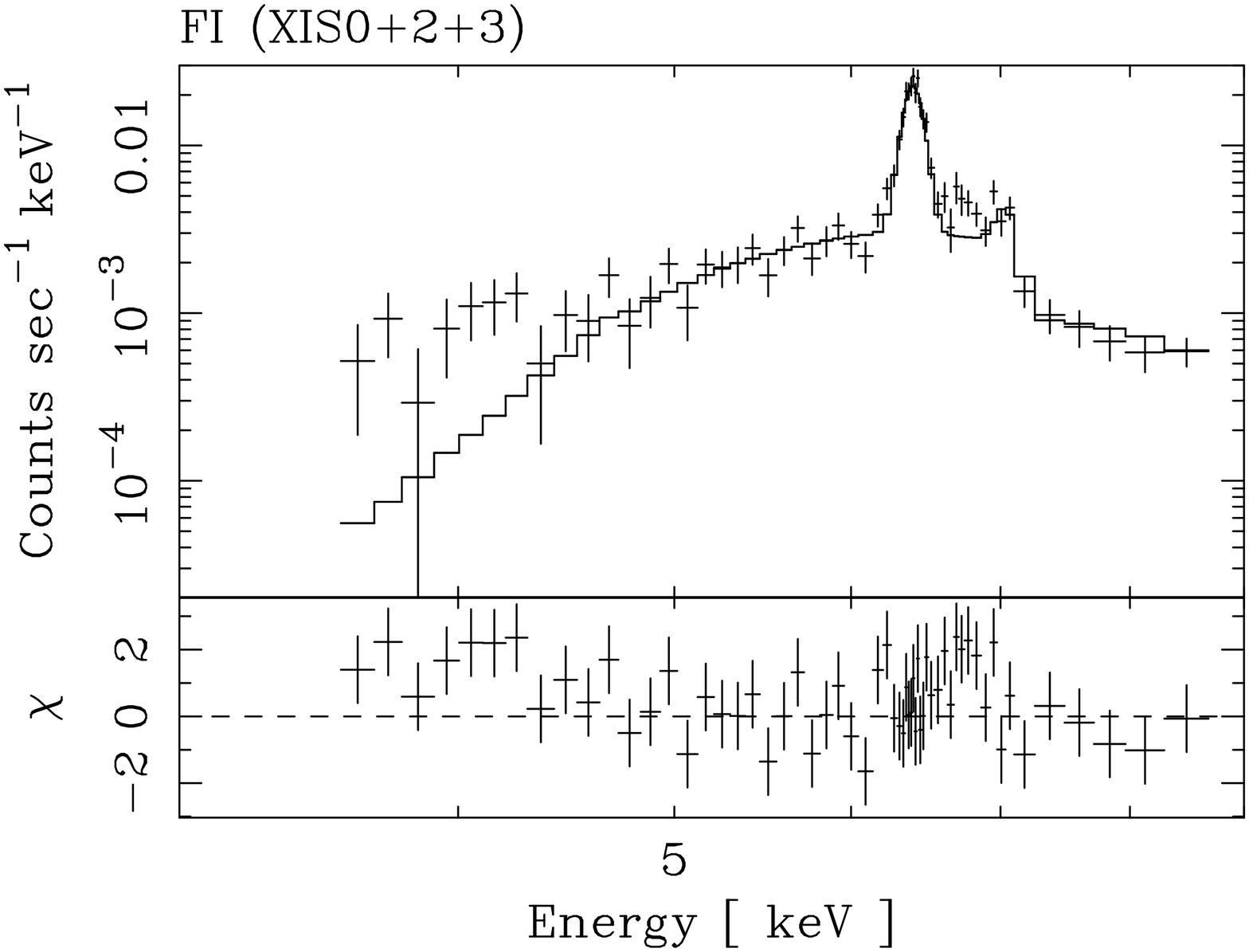}
    \FigureFile(80mm,80mm){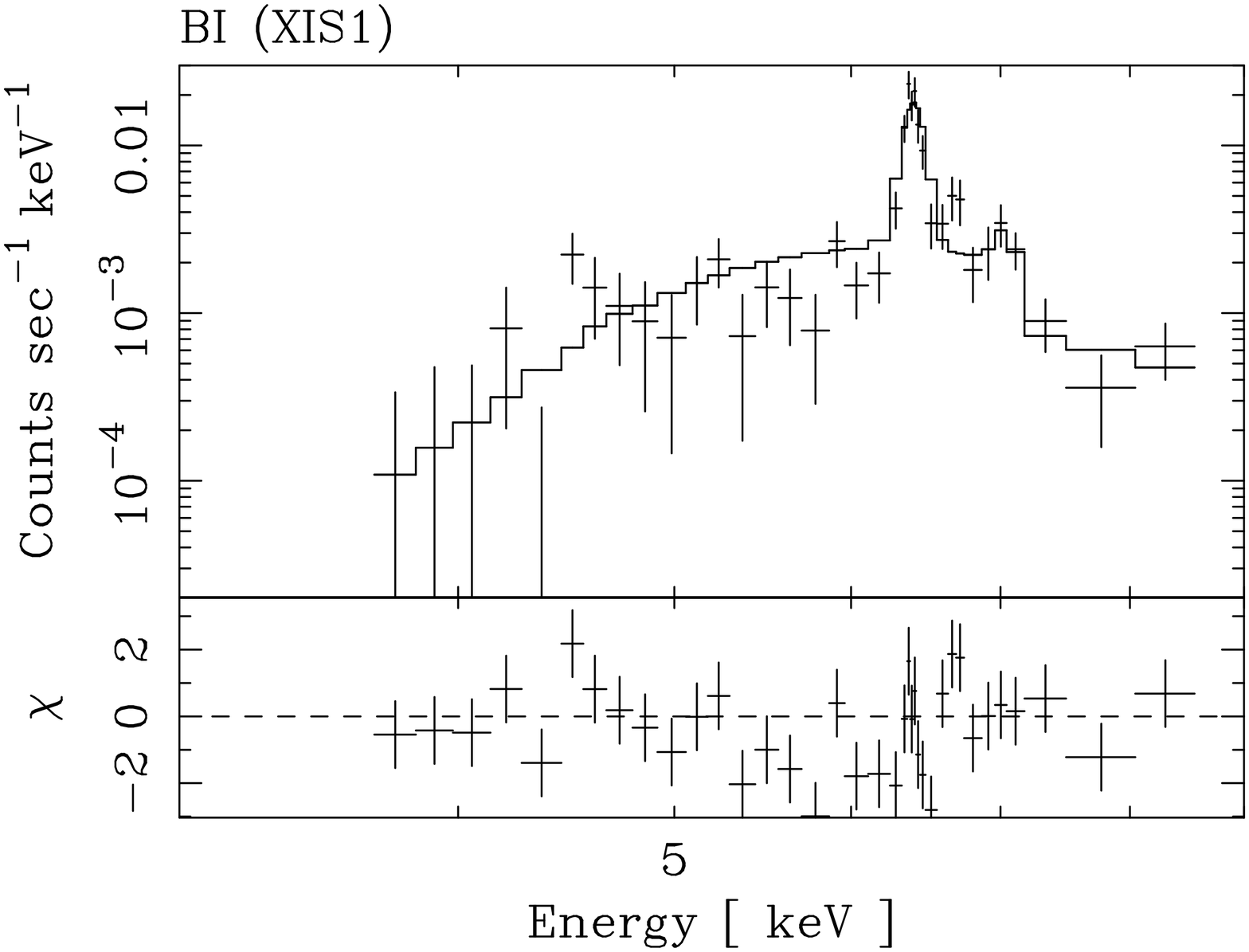}
  \end{center}
  \caption{Left: the X-ray spectrum of the sum of the 3 FI CCDs (XIS0, 2 and 3) for Sgr B2 with an absorbed power-law model and two Gaussian lines. Right: same as the right but of the BI CCD (XIS1)}
  \label{fig:sgrb-B2-spec}
\end{figure}

\begin{figure}
  \begin{center}
    \FigureFile(80mm,80mm){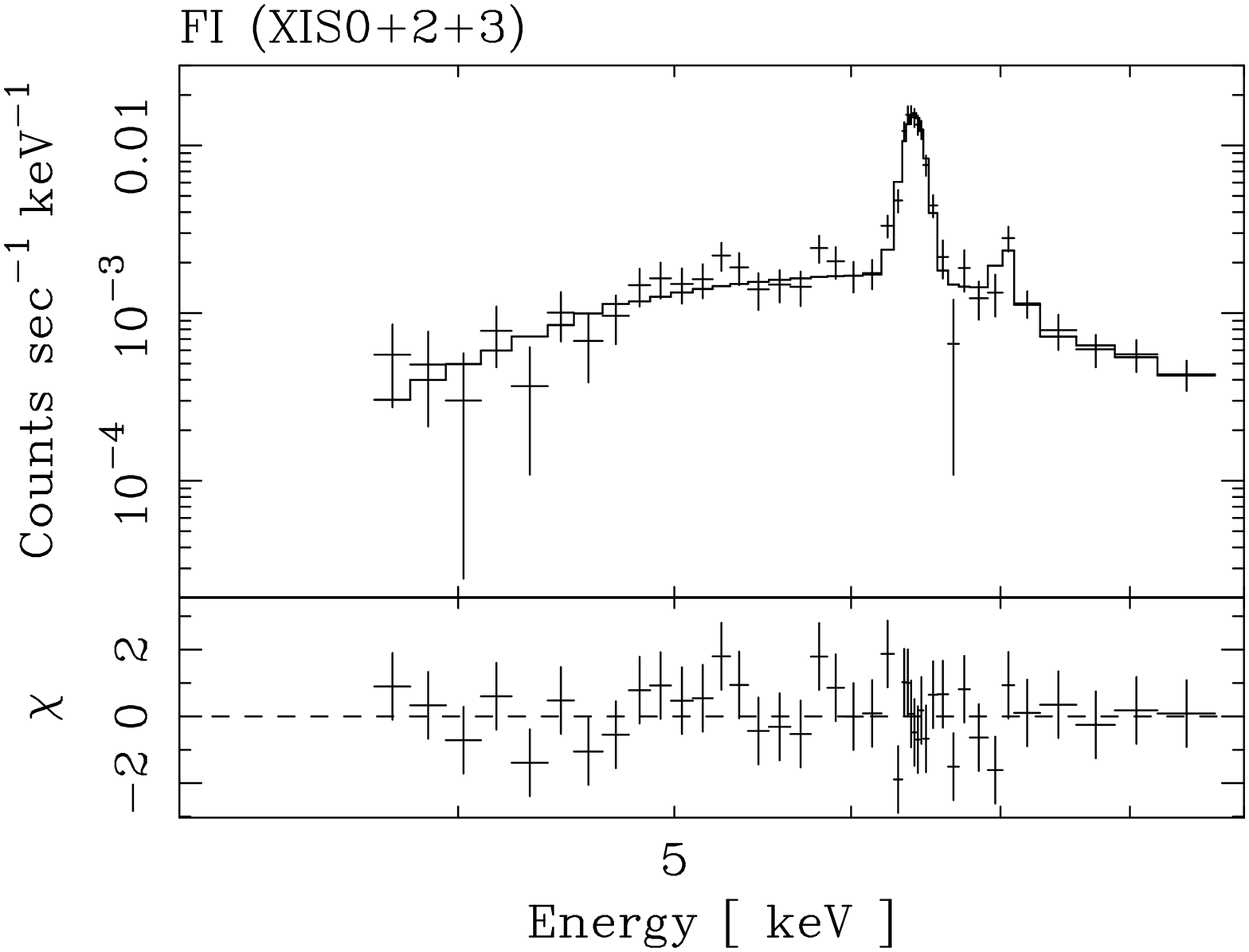}
    \FigureFile(80mm,80mm){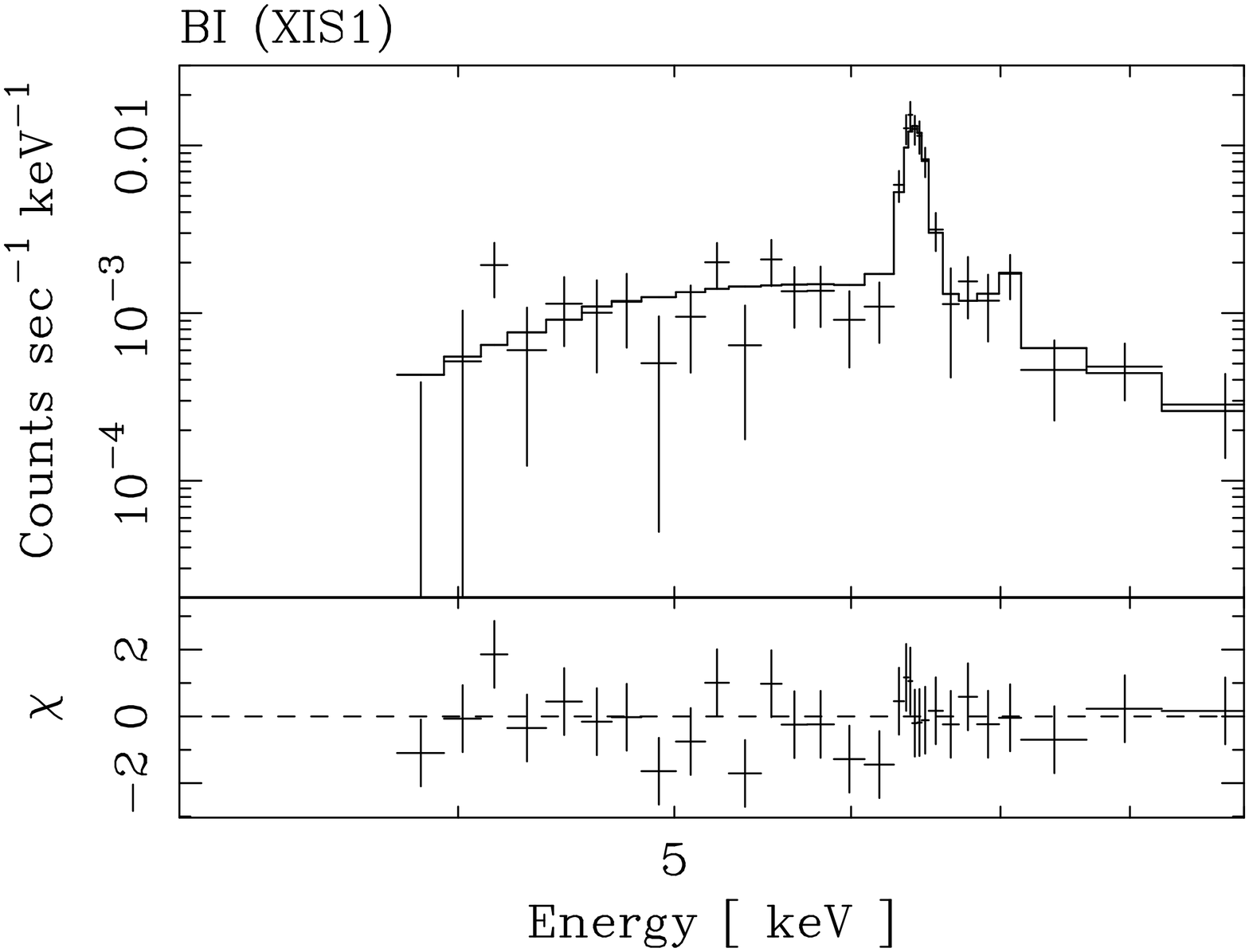}
  \end{center}
  \caption{Same as figure \ref{fig:sgrb-B2-spec}, but for a new source M0.74-0.09}
  \label{fig:sgrb-m074009-spec}
\end{figure}

The background-subtracted spectra are shown is figures \ref{fig:sgrb-B2-spec} and \ref{fig:sgrb-m074009-spec}. We simultaneously fit the FIs and BI spectra with a model of absorbed power-law plus two Gaussians near at 6.4 and 7.06~keV, which are for the K$\alpha$ and K$\beta$ lines of  Fe\emissiontype{I}. The best-fit parameters are shown in table \ref{tab:m_fit}.
This model nicely fits the data except an excess near the 6.7~keV line in the Sgr B2 spectra. In fact, the 6.7~keV line map (figure \ref{fig:sgrb-6700img}) shows a weak enhancement at the position of Sgr B2. One possibility is that the 6.7~keV enhancement is a part of the new SNR candidate G0.61+0.01, because it is located in the close vicinity of Sgr B2. The other possibility is that the 6.7~keV enhancement is due to YSOs embedded in the center of Sgr B2. In fact, the Sgr B2 region is 
relatively crowded  with the Chandra point sources (13 point  sources), and at least some of them are  YSOs  with  a hint of the 6.7~keV line emission \citep{Takagi2002}.  The total flux (in the 2--10 keV band) of the point sources is 
$\sim$10$^{-13}$~ergs~cm$^{-2}$~s$^{-1}$, which is $\sim$6\% of the Sgr B2 flux (see table 3).

The Sgr B2 cloud has been studied extensively with ASCA and Chandra.  \citet{Koyama1996} and \citet{Murakami2001} concluded that the 6.4~keV emission is due to  fluorescence by strong X-rays coming from Sgr A$^{*}$, hence named the X-ray reflection nebula (XRN). In this paper, we found a clear K$\beta$ line at 7.06~keV with consistent flux ratio to the K$\alpha$ line (6.4~keV) in the fluorescent X-ray origin and deep Fe edge at 7.1~keV. These discoveries provide additional supports for the XRN scenario of Sgr B2. Further details on the Sgr B2 results with Suzaku will be presented in a separate paper.

In the M0.74$-$0.09 region, Miyazaki and Tsuboi (2000) reported flux peaks of the CS (J=1-0) line emission at $(l, b) = (\timeform{0D.761}, -\timeform{0D.117})$ and $(\timeform{0D.764}, -\timeform{0D.064})$, clear evidence for the presence of a molecular cloud. The XIS spectrum of this region exhibits a strong 6.4~keV line with an equivalent width of 1.6~keV, a 7.06~keV line and an Fe edge structure at 7.1~keV (see table 3). All these features are consistent with being from K$\alpha$, K$\beta$ and K-edge from Fe\emissiontype{I}. The flux of the 7.06~keV line is about 10\% of that of the 6.4~keV line, which is also consistent with the fluorescent X-ray origin (Kaastra and Mewe 1993). We note here that the background region is the same as the case of G0.61+0.01, hence with the same argument in section 3.2, possible point source contribution can be ignored.

Unlike Sgr B2, no hint of HM YSO is found so far.  No bright point source is found in the Chandra image. 
Therefore, the X-rays can not be the scattering and fluorescence by embedded YSOs. If the X-rays from Sgr B2 and M0.74$-$0.09 are due to the Thomson scattering and fluorescence of the same irradiating external source like Sgr A$^*$, then the $N_{\rm H}$ ratio between these sources should be similar to the 6.4~keV line flux ratio. The observed $N_{\rm H}$ ratio is 0.42, while that of the 6.4~keV line flux is 0.36, in good agreement of the fluorescence scenario by a single irradiation source.  Therefore the XRN scenario by the past activity of Sgr A$^*$, which was successfully applied for Sgr B2 may also be applied for M0.74$-$0.09.

The counter scenario against the XRN is that the 6.4~keV line emission is produced by the collision of electrons. Since the cross section of iron K-shell ionization is maximum at the electron energy of a few 10~keV \citep{Tatischeff2002}, the most probable source is low energy electrons (LEE) as proposed for the origin of the Galactic Ridge iron K-shell emission \citep{Valinia2000}. Since a few 10~keV electrons are 
absorbed in less than $10^{22}$~H~cm$^{-2}$ of depth \citep{Tatischeff2002}
, the produced X-ray spectrum should have no large absorption edge. Our observation, however, shows a clear absorption of 
(4.0--9.6) $\times 10^{23}$~H~cm$^{-2}$, in far excess to the Galactic interstellar absorption \citep{Sakano2002}. Thus the LEE origin is unlikely, unless we assume a special geometry such that the 6. 4keV source is deep in or behind the dense cloud.   

\begin{table*}
  \caption{The result of spectral fittings of Sgr B2 and M0.74$-$0.09 with 
  a power-law and two Gaussian models}\label{tab:m_fit}
 \label{tab:m_fit}
  \begin{center}
    \begin{tabular}{lcc}
      \hline\hline
      Parameter & Sgr B2 & M0.74$-$0.09 \\
      \hline
      Absorbed power-law model: & & \\
      Column density $N_{\rm H}$ ($10^{23}$ cm$^{-2}$) & $9.6_{-0.8}^{+2.5}$ & $4.0_{-1.1}^{+1.4}$ \\
      Photon index $\Gamma$ & $3.2_{-0.6}^{+0.9}$ & $1.4_{-0.7}^{+0.4}$ \\
      Gaussian 1 (Fe\emissiontype{I} K\emissiontype{$\alpha$}): & & \\
      Line energy (eV) & $6399_{-5}^{+5}$ & $6406_{-6}^{+6}$ \\
      Intensity ($10^{-5}$ photons cm$^{-2}$ s$^{-1}$) & $16.5_{-0.3}^{+0.8}$ & $5.9_{-1.0}^{+1.4}$ \\
      Equivalent Width (keV) & 1.13 & 1.55 \\
      Gaussian 2 (Fe\emissiontype{I} K\emissiontype{$\beta$}): & & \\
      Line energy (eV)\footnotemark[$a$] & 7058 & 7065 \\
      Intensity ($10^{-5}$ photons cm$^{-2}$ s$^{-1}$) & $1.4_{-0.5}^{+0.5}$ & $0.6_{-0.3}^{+0.3}$ \\
      Equivalent Width (keV) & 0.13 & 0.18 \\
      \hline
      Observed flux\footnotemark[$\dagger$] ($10^{-12}$ ergs cm$^{-2}$ s$^{-1}$) & $1.5_{-0.9}^{+0.1}$ & $1.3_{-0.8}^{+0.2}$ \\
	  Luminosity\footnotemark[$\ddagger$] ($10^{34}$ ergs s$^{-1}$) & $9.7_{-5.1}^{+0.1}$ & $2.6_{-0.9}^{+0.4}$ \\
      $\chi^2$/dof & 154.7/89 & 54.4/66 \\
      \hline
      \multicolumn{3}{@{}l@{}}{\hbox to 0pt{\parbox{180mm}{\footnotesize
      	Note---The uncertainties indicate the 90\% confidence limit.
      	\par\noindent
      	\footnotemark[$a$] The energy gap between K$\alpha$ and K$\beta$ is fixed at the theoretical value (+659~eV) 
(Kaastra and Mewe 1993).
      	\par\noindent
      	\footnotemark[$\dagger$] Observed flux in the 4.0--10.0~keV band.
      	\par\noindent
      	\footnotemark[$\ddagger$] Absorption corrected luminosity in the 4.0--10.0~keV band.
      }\hss}}
    \end{tabular}
  \end{center}
\end{table*}

\section{Summary}
We summarize the results of the Sgr B observation as follows;
\begin{enumerate}	
\item All the Sgr B region is covered with a thin hot plasma, which is regarded as a part of the GCDX.
\item The Sgr B region is separately mapped with the 6.4~keV and 6.7~keV lines.
\item We found a local excess in the 6.7 keV line named as G0.61+0.01, which  is likely an ejecta dominant SNR.
\item The 6.4~keV map shows local excess at the giant molecular cloud Sgr B2 and M0.74$-$0.09. Like Sgr B2, M0.74$-$0.09 is a good candidate of an XRN.

\end{enumerate}

\bigskip
The authors  thank  all the Suzaku team members, especially  T. Takahashi, A. Senda, A. Bamba, 
J. Kataoka, Y. Tsuboi, H. Uchiyama, H. Nakajima, H. Yamaguchi, and H. Mori for their comments, supports and useful information on the XIS performance.
T.I. and H.Y. are supported by JSPS Research Fellowship for Young Scientists.
This work is supported by the Grant-in-Aid for the 21st Century COE "Center for Diversity and Universality in Physics" from the Ministry of Education, Culture, Sports, Science and Technology (MEXT) of Japan.

\end{document}